\documentclass[twocolumn,preprintnumbers,pra,smaller,
superscriptaddress]{revtex4}

\usepackage{amssymb,amsmath,graphicx,dcolumn,bm,float} 
\usepackage[utf8]{inputenc}
\usepackage[T1]{fontenc}
\usepackage[colorlinks,linkcolor=black,citecolor=blue]{hyperref}

\begin{document}

\title{Exceptional point enhances sensitivity of optomechanical mass sensors}
\author{P. Djorwe}
\email{philippe.djorwe@univ-lille1.fr}
\affiliation{Institut d’Electronique, de Microélectronique et Nanotechnologie, UMR
CNRS 8520 Université de Lille, Sciences et technologies, Villeneuve d’ Ascq 59652, France}

\author{Y. Pennec}
\email{yan.pennec@@univ-lille1.fr}
\affiliation{Institut d’Electronique, de Microélectronique et Nanotechnologie, UMR
CNRS 8520 Université de Lille, Sciences et technologies, Villeneuve d’ Ascq 59652, France}

\author{B. Djafari-Rouhani}
\email{bahram.djafari-rouhani@@univ-lille1.fr}
\affiliation{Institut d’Electronique, de Microélectronique et Nanotechnologie, UMR
CNRS 8520 Université de Lille, Sciences et technologies, Villeneuve d’ Ascq 59652, France}

\begin{abstract}
We propose an efficient optomechanical mass sensor operating at exceptional points (EPs),
non-hermitian degeneracies where eigenvalues of a system and their corresponding 
eigenvectors simultaneously coalesce. The benchmark system consists of two optomechanical cavities 
(OMCs) that are mechanically coupled, where we engineer mechanical gain (loss) by driving the cavity 
with a blue (red) detuned laser. The system features EP at the gain and loss balance, where 
any perturbation induces a frequency splitting that scales as the square-root of the 
perturbation strength, resulting in a giant sensitivity factor enhancement compared to the conventional optomechanical 
sensors. For non-degenerated mechanical resonators, quadratic optomechanical coupling is used to tune the mismatch 
frequency in order to get closer to the EP, extending the efficiency of our sensing scheme to mismatched resonators. 
This work paves the way towards new levels of sensitivity for optomechanical sensors, which could find 
applications in many other fields including nanoparticles detection, precision measurement, and quantum metrology.
\end{abstract}

\maketitle

\date{\today}


%
\section{Introduction} \label{Intro}

Owing to a rapid progress in micro/nano-engineering of mechanical resonators, tremendous advances have been 
made in mass sensing in the view of achieving ultra sensitive detection \cite{[1],[2]}.
Resonant mass sensors are widely used in diverse fields of science and technologies. For instance, they are 
actively used to detect single biomolecules, viruses, nanoparticles and others nano-objects. Because of these 
interesting applications, nano-mechanical mass sensors have deserved much attention 
over the last few years. This has led to detections ranging from 
femtogram ($10^{-15}\rm{g}$) \cite{[3]} though attogram ($10^{-18}\rm{g}$) \cite{[4]} and zeptogram 
($10^{-21}\rm{g}$) \cite{[5]} to yoctogram ($10^{-24}\rm{g}$) \cite{[6]}.

Besides these sensors based on electrical excitations, all-optical mass sensors have been proposed as well, 
with the aim of reaching new levels of sensitivity and to breakthrough the limitation of frequency restriction 
(see \cite{[7]} and the references therein). Optomechanical sensors, have been revealed as excellent 
candidates for mass detection due to their simplicity, sensitivity and because the optical field 
serves both as an actuator and a probe for precise monitoring of the mechanical frequencies. 
Single optomechanical sensors, exploiting nonlinearies of the systems, 
were proposed in \cite{[8],[9],[10]} whereas coupled optomechanical cavities have been used for 
sensing in \cite{[11],[12]}. A mechanical resonator was inserted inside one of the coupled optical cavities in 
\cite{[11]}, while a hybrid opto-electromechanical system consisting of an optical and microwave cavities coupled
to a common mechanical oscillator has been used in \cite{[12]}. It resulted that sensors based on coupled 
systems perform better than their counterpart that are based on single cavity. 
This conclusion remains the same even for coupled mechanical resonators 
driving by a common electromagnetic field as shown in \cite{[13],[14]}. The main reason of such performance lies on the 
fact that, sensitivity in coupled structures is determined through the splitting of eigenmodes, and it 
therefore can be significantly increased even without entering the strong coupling regime. This is not the case neither for 
electromechanical nor single optomechanical mass sensors aforementioned, where the sensing process employs tracking the frequency 
shifts of the mechanical resonator due to mass changes induced by any added tiny object \cite{[1],[2]}. Moreover, simple multimodal 
response of a mechanical resonator enables mass sensing at much higher levels of accuracy than what is typically 
achieved with a single frequency shift measurement \cite{[15]}. It goes without saying that mass sensor performances 
are enhanced in coupled systems that feature strong coupling regime or those exhibiting very narrow splitting
in their eigenmodes structure. 

Recently, sensitivity enhancement of more than threefold in particles detection has been theoretically demonstrated with 
a sensor operating at exceptional points (EPs) \cite{[16]}. Owing to the complex square-root topology near an EP, 
any perturbation lifts the degeneracy leading to 
a frequency splitting that scales as the square-root of the perturbation strength. This splitting 
is therefore larger for sufficiently small perturbation than the splitting observed in conventional  
sensing schemes, where the linewidth or frequency shift/splitting is proportional 
to the strength of the perturbation. Later on, these theoretical results have been validated by two experimental 
works, at EP in whispering-gallery-mode micro-toroid cavity \cite{[17]} and with higher order EP in parity-time 
($\mathcal{PT}$)-symmetric photonic laser molecule \cite{[18]}. Along the same line, the performance of cavity-assisted 
metrology, where a cavity is coupled to the device under test, has been enhanced 
near EP in $\mathcal{PT}$-symmetric micro-cavities \cite{[19]}. It follows that EP 
is a useful tool for next generation of sensors in reaching new levels of sensitivity. 

Our proposal is based on two optomechanical cavities, mechanically coupled through their moving mechanical resonators.
This benchmark system has an advantage that it enables to get an effective coupled mechanical system, where mechanical gain (loss) can 
be easily engineered by driving the cavity with a blue (red) detuned laser. At the balance between gain and loss, the system features EP in its mechanical 
spectrum \cite{[20]}, which is the key point of our optomechanical mass sensor-based EP. The proposed sensor differs from 
those known \cite{[16],[17],[18],[19]} on at least three points: (i) it operates at an 
EP generated on its mechanical spectrum (instead of the optical spectrum), (ii) the gain and loss are self-engineered 
while the cavities are driven, and (iii) it does not needs the $\mathcal{PT}$-symmetric requirement. At the vicinity 
of EP, our proposed sensing scheme has shown a giant enhancement sensitivity factor compared to conventional 
opto-(electro)mechanical sensing schemes. This efficiency comes from topological complex square-root near the EP, 
and is highly pronouced for small perturbations. Bearing in mind the laborious micro/nanofabrication task of engineering 
two identical mechanical resonators, we have extended our results to non-degenerated mechanical resonators, where 
their frequency mismatch can be tuned through quadratic coupling that is well known in optomechanics \cite{[21]}. 
This work is organized as follows. In Sec. \ref{MoEq}, the model and its dynamical equations up to the EP features
are described. The sensitivity enhancement, at the vicinity of EP for degenerated mechanical resonators, 
is presented in Sec. \ref{Sen.D}. Section \ref{Sen.N.D} is devoted to 
the sensitivity assisted by quadratic coupling for the non-degenerated case, while Sec. \ref{Concl} concludes the work.

\section{Modelling and dynamical equations} \label{MoEq}

\begin{figure}[tbh]
\begin{center}
\resizebox{0.4\textwidth}{!}{
\includegraphics{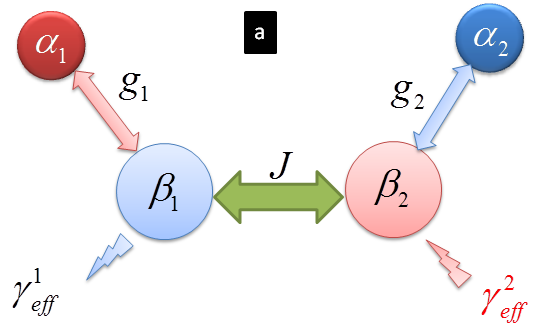}}
\resizebox{0.4\textwidth}{!}{
\includegraphics{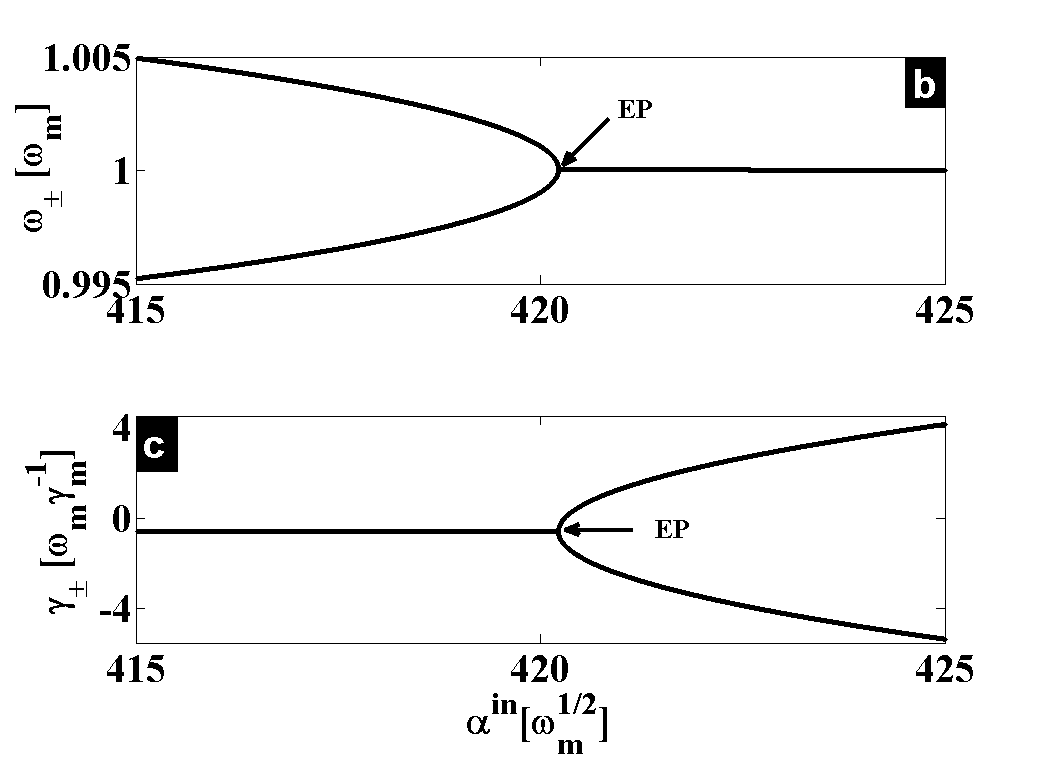}}
\end{center}
\caption{(a) Generic setup. Two optomechanical cavities, one with gain and the other generating 
losses are mechanically coupled. (b)-(c) Real and imaginary parts of the eigenvalues versus the driving strength $\alpha^{in}$, 
which depict the EP feature.}
\label{fig:Fig1}
\end{figure}

The system of our proposal is the one sketched in Fig. \ref{fig:Fig1}a. 
It consists of two optomechanical cavities, mechanically coupled through their moving mechanical resonators. 
The first cavity is driven with a red-detuned laser while the second one is excited with a blue-detuned driving field. 
By symmetrically driving the cavities at the mentioned sidebands, we can engineer either mechanical gain or loss. 
In the rotating frame of the driving fields ($\omega_{p}$), 
the Hamiltonian ($\hbar=1$) describing this system is,
\begin{equation}
H=H_{OM}+H_{int}+H_{drive}, \label{eq1}
\end{equation}
with 
\begin{equation}
\left\{
\begin{array}
[c]{c}
H_{OM}=\sum_{j=1,2}\omega_{j}b_{j}^{\dag}b_{j}-\Delta_{j} a_{j}^{\dag}a_{j}\\
-ga_{j}^{\dag}a_{j}(b_{j}^{\dag}+b_{j}) \\
H_{int}=-J(b_{1}b_{2}^{\dag}+b_{1}^{\dag}b_{2}) \\
H_{drive}=E(a^{\dag}+a).
\end{array}
\right. \label{eq2}
\end{equation}
In this Hamiltonian, $a_{j}$ and $b_{j}$ are the annihilation  bosonic field operators describing the optical and mechanical resonators, 
respectively. The mechanical displacements $x_{j}$ are connected to $b_{j}$ as $x_{j}=x_{_{\rm{ZPF}}}(b_{j} +b_{j}^{\dag})$, 
where $x_{_{\rm{ZPF}}}$ is the zero-point fluctuation amplitude of the mechanical resonator. The mechanical frequency 
of the $j^{th}$ resonator is $\omega_{j}$ and  $\Delta_{j}=\omega_{p}^{j}-\omega_{cav}^{j}$ is the
optical detuning between the  optical and the  cavity ($\omega_{cav}^{j}$) frequencies . 
The mechanical coupling strength between the two mechanical resonators is $J$, and  the linear optomechanical  
coupling is $g$. 
By setting the mean values of the operators, $\langle a \rangle=\alpha_{j}$ for the optics 
and $\langle b \rangle=\beta_{j}$ for the mechanics, one can derive the following classical set of nonlinear equations
of our  system,  
\begin{equation}
\left\{
\begin{array}{c}
\dot{\alpha}_{j}=[i(\Delta_{j}+g(\beta_{j}^{\ast}+\beta
_{j})) -\frac{\kappa }{2}] \alpha_{j}-i\sqrt{\kappa}\alpha^{\rm{in}}, \\
\dot{\beta}_{j}=-(i\omega_{j}+\frac{\gamma_{m}}{2}) \beta_{j}+iJ\beta_{3-j}+ig\alpha_{j}^{\ast}\alpha_{j},
\end{array}
\right.  \label{eq3}
\end{equation}
where optical ($\kappa$) and mechanical ($\gamma$) dissipations have been added, and the amplitude of the driving 
pump has been substituted as $E=\sqrt{\kappa}\alpha^{\rm{in}}$ in order to account for losses. In this form, the input laser power 
$\rm{P_{in}}$ acts through $\alpha^{\rm{in}}=\sqrt{\frac{\rm{P_{in}}}{\hbar\omega_{p}}}$. 
Without loss of generality, the parameters $\gamma_{m}$, $g$, and $\kappa$ 
are assumed to be degenerated for the whole system. Moreover, they follows the hierarchy
$\gamma_{m},g\ll \kappa \ll \omega_{m}$, as encountered in resolved sideband regime experiments \cite{[22],[23]}.

To investigate sensitivity performance of our proposed sensor, 
we need to compute the eigenvalues of the effective mechanical system in order to identify 
the EP feature. One can obtain this effective system by integrating $\alpha_{j} (t)$ 
out of the full set of Eq. (\ref{eq3}). This can 
be straightforwardly done by approaching the mechanical oscillations with the ansatz, 
$\beta_{j}(t)=\bar{\beta}_{j}+A_{j}\exp(-i\omega_{_{\rm{lock}}}t)$ (see more details in \cite{[20]}). 
$\bar{\beta}_{j}$ is a constant shift in the origin of the movement,
 $A_{j}$ is the slowly time dependent amplitude, and $\omega_{_{\rm{lock}}}$ is the 
 mechanical degenerated frequency when the resonators experience frequency locking phenomenon.
This leads to the following mechanical effective Hamiltonian,

\begin{equation}
H_{\rm{eff}}=
\begin{bmatrix}
\omega_{\rm{eff}}^{1}-i\frac{\gamma_{\rm{eff}}^{1}}{2} & -J \\
-J & \omega_{\rm{eff}}^{2}-i\frac{\gamma_{\rm{eff}}^{2}}{2}
\end{bmatrix},
\label{eq4}
\end{equation}
having the eigenvalues

\begin{equation}
\lambda_{\pm}=  \frac{\omega_{\rm{eff}}^{1}+\omega_{\rm{eff}}^{2}}{2}-\frac{i}{4}\left(\gamma_{\rm{eff}}^{1}+\gamma
_{\rm{eff}}^{2}\right) \pm \frac{\sigma}{4},  \label{eq5}
\end{equation}
with 
\begin{equation}
\sigma = \sqrt{16J^{2}+[2(\omega_{eff}^{1}-\omega_{eff}^{2}) 
+i(\gamma_{eff}^{2}-\gamma_{eff}^{1})]^2}. \label{eq6}
\end{equation}
Here, $\omega_{\rm{eff}}^{j}=\omega_{m}+\delta \omega_{opt}^{j}$
and $\gamma_{\rm{eff}}^{j}=\gamma_{m}+\gamma_{\rm{opt}}^{j}$ are the effective frequencies
and  dampings, respectively. Furthermore, the quantities $\delta \omega_{opt}^{j}$ and $\gamma_{\rm{opt}}^{j}$ 
are the optical spring effect and optical damping induced by the optical fields, respectively. These terms 
are given as, 

\begin{equation}
\delta \omega_{opt}^{j}=-\frac{2\kappa (g\alpha^{in})^{2}}{\omega
_{lock}\epsilon_{j}}\rm{Re}\left(\sum_{n}\frac{J_{n+1}\left(
-\epsilon_{j}\right) J_{n}\left(-\epsilon_{j}\right)}{
h_{n+1}^{j\ast}h_{n}^{j}}\right),  \label{eq7}
\end{equation}

and 

\begin{equation}
\gamma_{\rm{opt}}^{j}=\frac{2(g\kappa \alpha ^{\rm{in}})^{2}}{\epsilon_{j}}
\sum_{n}\frac{J_{n+1}\left(-\epsilon_{j}\right) J_{n}\left(
-\epsilon_{j}\right)}{\left\vert h_{n+1}^{j\ast}h_{n}^{j}\right\vert
^{2}},  \label{eq8}
\end{equation} 
where $\epsilon_{j}=\frac{2g\rm{Re}(A_{j})}{\omega_{_{\rm{lock}}}}$ is a normalized amplitude,  
$J_{n}$ the Bessel function,  $h_{n}^{j}=i\left(n\omega_{_{\rm{lock}}}-\tilde{\Delta}_{j}\right) 
+\frac{\kappa}{2}$, and $\tilde{\Delta}_{j}=\Delta_{j}+2g(\bar{\beta}_{j})$ 
the effective detuning \cite{[20]}. 
The eigenfrequencies and the dampings of the system are
defined as the real [$\omega_{\pm}=\rm{Re}(\lambda_{\pm})$] and imaginary [$\gamma_{\pm}=\rm{Im}(\lambda_{\pm})$]
parts  of $\lambda_{\pm}$, respectively.

\begin{figure}[tbh]
\begin{center}
\resizebox{0.4\textwidth}{!}{
\includegraphics{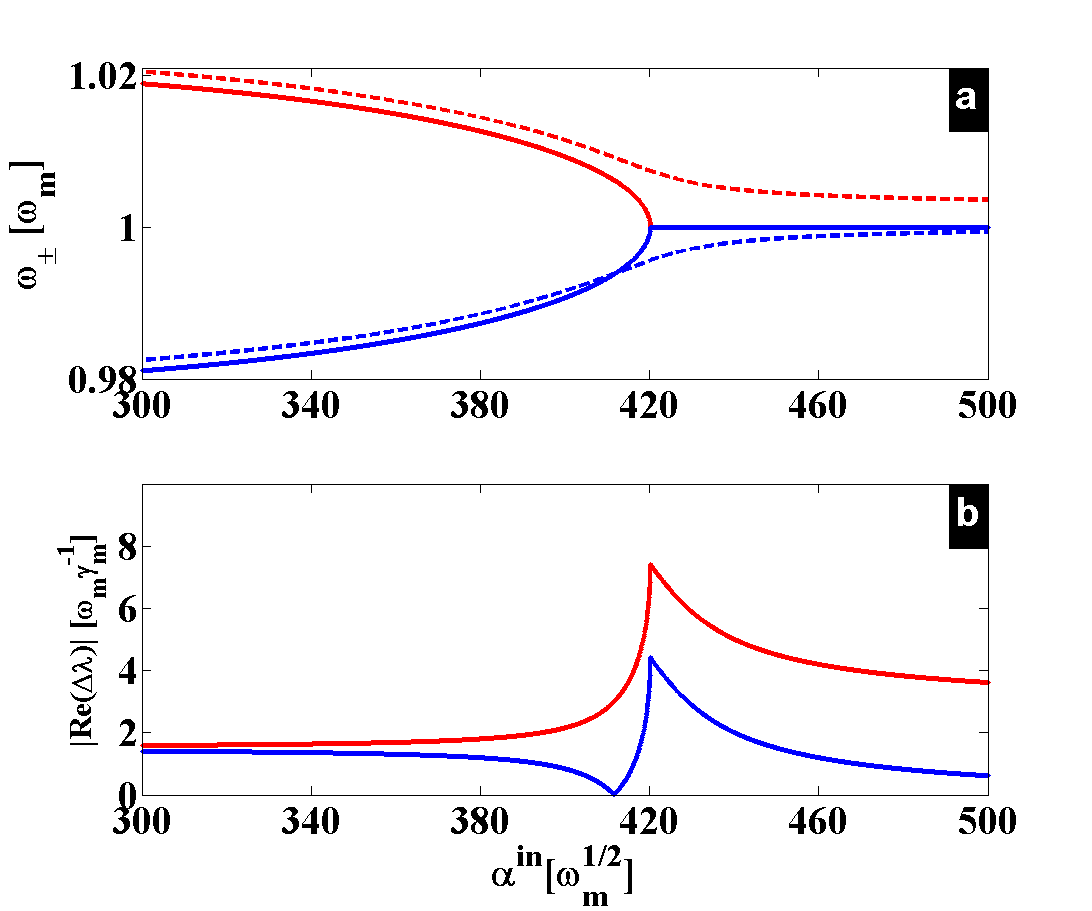}}
\end{center}
\caption{(a) Frequencies of the effective mechanical system, before perturbation (full lines) and after perturbation 
by a mass deposition (dashed lines). (b) Gap difference between the perturbed and reference frequencies shown in (a). These 
quantities are plotted versus the driving strength $\alpha^{in}$. 
Note the difference in the units on the $y$-axis in (a) and (b).}
\label{fig:Fig2}
\end{figure}

At the EP, both these pairs of frequencies and dampings coalesce, $\omega_{-}=\omega_{+}\equiv\omega_{lock}$ and 
$\gamma_{-}=\gamma_{+}$. From Eq. (\ref{eq5}), this is equivalent to $\sigma=0$, which requires both control of real and 
imaginary parts of $\sigma$, through the driving fields ($\alpha^{in}_{j}$) and detunings ($\Delta_{j}$) \cite{[24]}. 
For a proof of concept of the proposed sensor, we assume in the next section that 
the two mechanical resonators are degenerated ($\omega_{j}\equiv\omega_{m}$). However, the non-degenerated 
case, where there is a mismatch frequency due to micro/nanofabrication imperfections for intance, 
is later on discussed in section Sec. \ref{Sen.N.D}. 

\section{Sensitivity at the exceptional point} \label{Sen.D}
\begin{figure}[tbh]
\begin{center}
\resizebox{0.4\textwidth}{!}{
\includegraphics{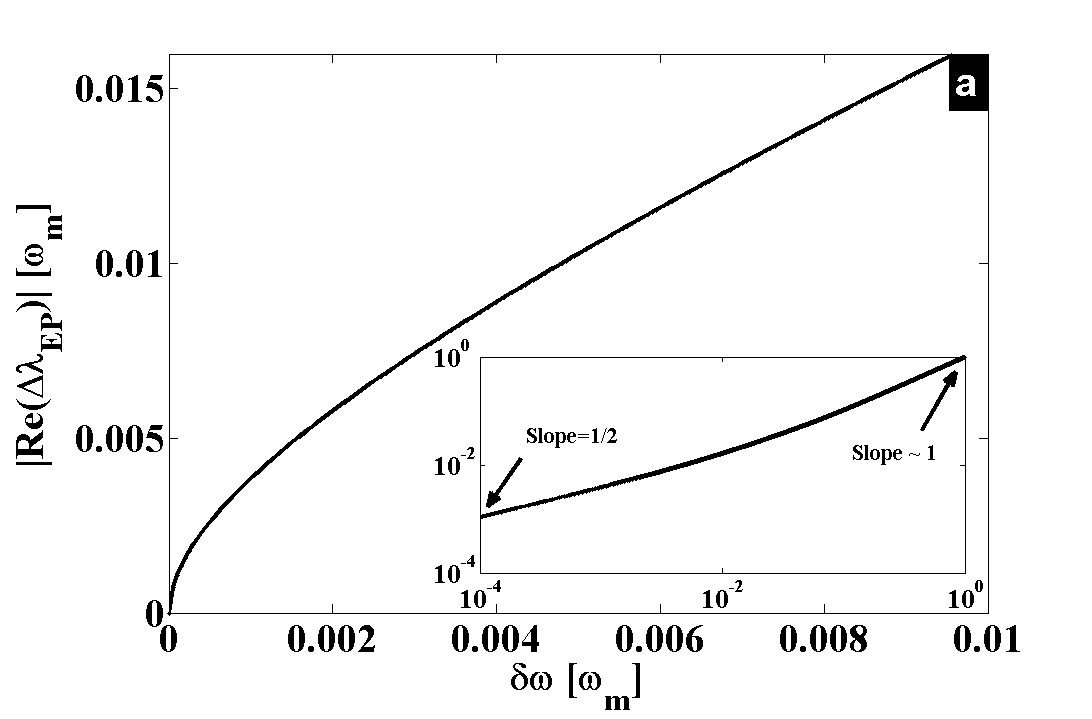}}
\resizebox{0.4\textwidth}{!}{
\includegraphics{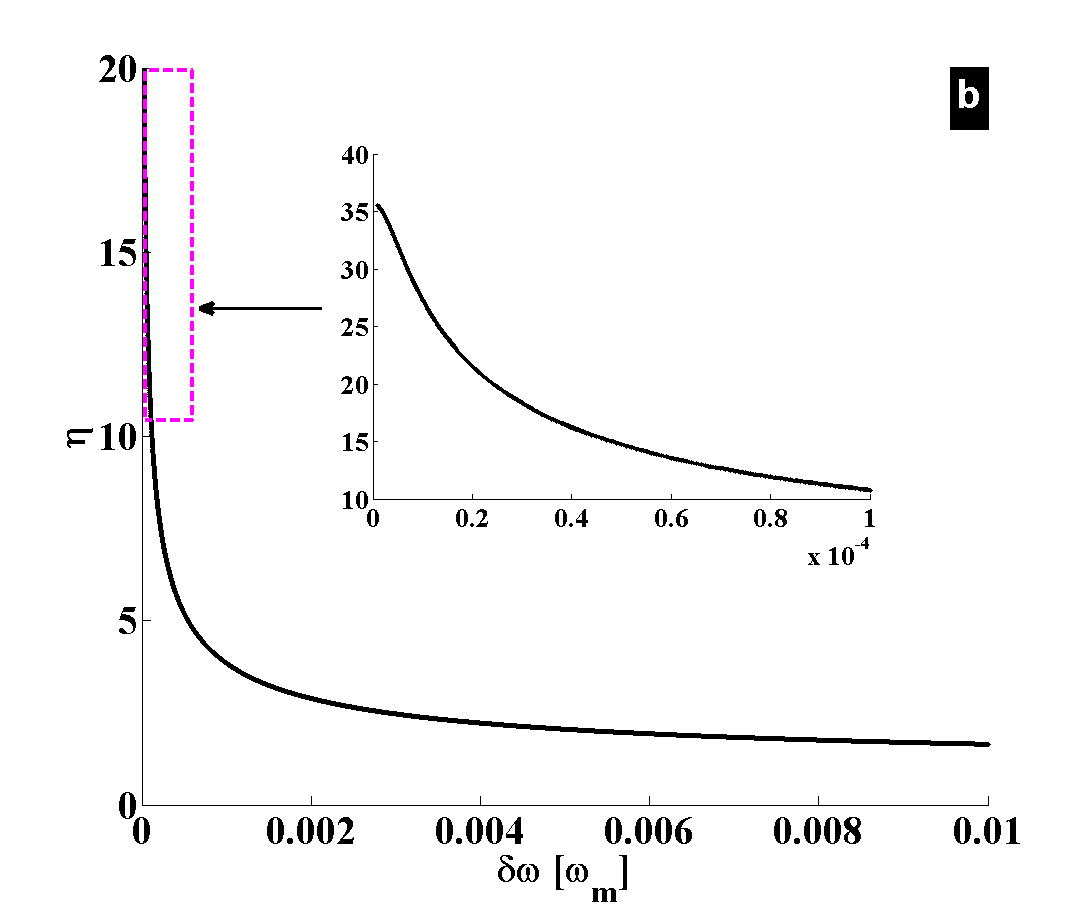}}
\end{center}
\caption{(a) Sensitivity at the exceptional point versus the strength of the perturbation  $\delta\omega$. 
The inset shows the log-log scale representation. (b) Sensitivity enhancement factor $\eta$ versus $\delta\omega$. 
It can be seen that for weak perturbation (see inset), the EP-sensor performs better, 
and ends up towards the same performance as the conventional ones as the perturbation increases.}
\label{fig:Fig3}
\end{figure}

The eigenvalues of our effective mechanical system undergo a singularity (known as EP) in their spectrum 
as shown in Fig. \ref{fig:Fig1}(b)-(c). The efficiency of the proposed sensor depends on this EP, which 
emerges from the eigenvalues given in Eq. (\ref{eq5}), instead of tracking the resonant frequency 
shift induced by mass deposition on the surface of the resonators \cite{[8],[9],[10],[11],[12],[13]}. 
For ordinary mass sensor, the relationship between the frequency shift $\delta\omega$ with the deposited 
mass $\delta m$ is given by \cite{[1],[2]},
\begin{equation}
\delta m=\frac{2m}{\omega_m}\delta\omega= \mathcal{R}^{-1}\delta\omega, \label{eq9}
\end{equation}
where $\mathcal{R}=\tfrac{\omega_m}{2m}$ stands for the mass responsivity and $m$ is 
the mass of the resonator supporting the deposition. It results that the induced 
frequency shift is proportional to the strength of the perturbation.  At the EP, both frequencies ($\omega_{\pm}$) 
and dampings ($\gamma_{\pm}$) coalesce (see Fig. \ref{fig:Fig1}b), and any external perturbation will split these 
eigenvalues leading to high sensitivity (see Fig. \ref{fig:Fig3}). To get insight of this feature, we assume that 
a mass deposition has landed on the mechanical resonator driven by the 
blue-detuned field ($m_{2}\equiv m$), which will induce its resonance to shift according to Eq. (\ref{eq9}). 
As the resonators are coupled, this localized shift will affect the whole system, resulting 
in a splitting at the EP. For degenerated resonators, and 
for weak driving regime one has $\delta \omega_{opt}^{j}\ll\omega_{m}$, and Eq. (\ref{eq5}) simplifies to 

\begin{equation}
\lambda_{\pm}\approx \omega_{m} -\frac{i}{4}\left(\gamma_{\rm{eff}}^{1}+\gamma
_{\rm{eff}}^{2}\right) \pm \sqrt{J^{2}-\left(\frac{\Delta\gamma_{eff}}{4}\right)^{2}},  \label{eq10}
\end{equation}
with $\Delta\gamma_{eff}=\gamma_{eff}^{2}-\gamma_{eff}^{1}$. As the second resonator 
experiences a small perturbation $\delta m$ (or shift $\delta\omega$), this affects 
the eigenvalues as (see appendix \ref{App.A}), 

\begin{equation}
\begin{split}
\lambda_{\pm}^{\delta\omega} & \approx \omega_{m} +\frac{\delta\omega}{2} -\frac{i}{4}\left(\gamma_{\rm{eff}}^{1}+\gamma
_{\rm{eff}}^{2}\right)\\
&\pm \sqrt{J^{2}-\left(\frac{\Delta\gamma_{eff}}{4}\right)^{2}
+\frac{\delta\omega^{2}+i\delta\omega\Delta\gamma_{eff}}{4}}.  \label{eq11}
\end{split}
\end{equation}
This perturbation generally affects also the optical dampings, but for the sake of a qualitative 
discussion and since the corresponding variations are very small (see Fig. \ref{fig:Fig6}(a)-(b) in appendix \ref{App.A}), we did not include them in Eq. (\ref{eq11}). 
However, it is worth noticing that these effects are fully accounted in our simulated results. 
We define sensitivity as the frequency shift of a supermode relative to its reference signal, i.e., 
the frequency splitting of these two pairs of supermodes 
$\rm{Re}(\Delta\lambda_{+})\equiv\rm{Re}(\lambda_{+}^{\delta\omega}-\lambda_{+})$ and 
$\rm{Re}(\Delta\lambda_{-})\equiv\rm{Re}(\lambda_{-}^{\delta\omega}-\lambda_{-})$ as 
shown in Fig. \ref{fig:Fig2}(a)-(b). One remarks that significant splitting is experienced at the EP 
(see Fig. \ref{fig:Fig2}(a)), resulting in a high sensitivity $|\rm{Re}(\Delta\lambda)|$ 
as depicted in Fig. \ref{fig:Fig2}(b). This feature can be approaching through our analytical 
investigation aforementioned. Indeed, the system undergoes the EP when 
$4J=\Delta\gamma_{eff}$ (see Eq. (\ref{eq10})). At this degeneracy point, and from Eq. (\ref{eq10}) and Eq. (\ref{eq11}), 
the splitting of the two pairs of supermodes can be deduced as (see appendix \ref{App.A}),
\begin{equation}
\Delta\lambda^{EP}=\frac{1}{2} \left(\delta\omega \pm \sqrt{\delta\omega^{2}-i\delta\omega\Delta\gamma_{eff}} \right).  \label{eq12}
\end{equation}
As expected, Eq. (\ref{eq12}) shows that there is no splitting at the EP for $\delta\omega=0$. 
For any perturbation however ($\delta\omega\neq0$), it reveals that the two 
pairs of supermodes experience different splittings $\rm{Re}(\Delta\lambda_{+})
\neq \rm{Re}(\Delta\lambda_{-})$, as it appears in Fig. \ref{fig:Fig2}(b). This asymmetry depends on 
which resonator the extra mass-like perturbation has landed. By adding the deposition on the other resonator instead, 
we have verified that the sensitivities of the supermodes are reversed. For small enough landed mass  
($|\delta\omega|\ll|\Delta\gamma_{eff}|$), both pairs of supermodes experience the same amount of 
splitting $\sim\sqrt{\tfrac{\delta \omega\Delta\gamma_{eff}}{8}}$ (see appendix \ref{App.A}). It results that the splitting at the 
EP scales as the square-root of the strength of the perturbation $\sim\delta\omega^{1/2}$, in stark 
contrast with the linear dependence as it is for the conventional sensors (see Eq. (\ref{eq9})). 
Owing to this complex square-root topology near the EP, the EP-sensors perform better in detecting 
small mass deposition compared to the conventional ones. In this weak perturbation 
limit, the splitting at the EP can be expressed in terms of the deposited mass as 
$\sim\sqrt{\frac{\omega_{m} \delta m \Delta\gamma_{eff}}{16m}}$. From this expression and Eq. (\ref{eq9}),
we can define the sensitivity enhancement factor as,

\begin{equation}
\eta\equiv\left|\frac{\rm{Re}(\Delta\lambda^{\rm{EP}})}{\delta\omega}\right|=\sqrt{\frac{\Delta\gamma_{eff}}{8\delta \omega}}
=\sqrt{\frac{m\Delta\gamma_{eff}}{4\omega_{m} \delta m}}.  \label{eqa10}
\end{equation}
which again highlights the square root dependence with $\delta m$, instead of a linear dependence, 
for small mass deposition. As expected, Fig. \ref{fig:Fig3}(a) shows the square-root-like evolution of the sensitivity 
versus perturbation ($\left|\rm{Re}(\Delta\lambda^{EP})\right|\propto \delta\omega^{1/2}$), which is revealed 
through the log-log scale representation in the inset. Hence, a $\frac{1}{2}$-slope is shown 
for weak perturbations, which evolves towards $1$-slope behaviour as the perturbation increases. 
This simply reveals the fact that, the sensitivity of EP-sensor scales linearly with the strength of the 
perturbations when they are large enough. This is confirmed through the enhancement factor $\eta$ depicted 
in Fig. \ref{fig:Fig3}(b), where it evolves towards the limit $\eta\sim1$ for strong enough perturbation. 
It can be also seen that this factor $\eta$ is highly enhanced for weak perturbation strength (see inset), 
proving the efficiency of the EP-sensor in detecting small particles, mass or objects. 

\begin{figure}[tbh]
\begin{center}
\resizebox{0.4\textwidth}{!}{
\includegraphics{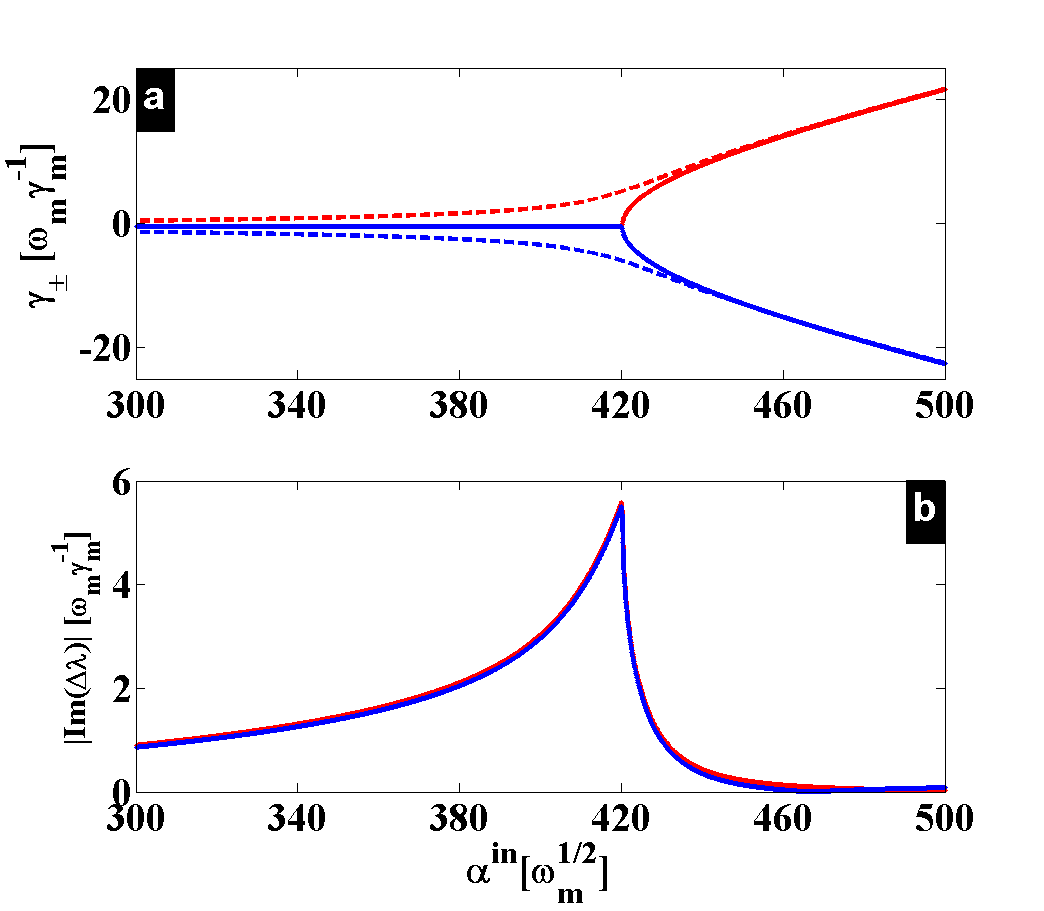}}
\end{center}
\caption{(a) Dampings as reference (full lines) and perturbed (dashed lines), with the 
associated splittings in (b). These plots are versus the driving strength $\alpha^{in}$.}
\label{fig:Fig4}
\end{figure}

Another interesting feature of EP-sensors is that they improve their performance regarding their linewidth as well. 
As the linewidths are also degenerated at the EP, adding extra mass will also 
lift this degeneracy. This is depicted in Fig. \ref{fig:Fig4}(a), and the corresponding sensitivity 
is shown in Fig. \ref{fig:Fig4}(b). In contrast to Fig. \ref{fig:Fig2}(b), here both sensitivities 
given by the pairs of supermodes are the same. This can be 
expected by refering to the imaginary part of Eq. (\ref{eq12}) as given in appendix \ref{App.A}. By further driving the system, the linewidths 
move out of the vicinity of EP, and one of them grows whereas the other decays. 
This simply means that one supermode experiences gain, while the other feels loss. 

\section{Enhancing sensitivity through quadratic coupling} \label{Sen.N.D}
The previous section has dealt with optomechanical sensor-based EP, where the coupled mechanical resonators are 
degenerated. However, engineering such identical resonators is not often an obious task in micro/nano-fabrication technologies. 
Therefore, any mismatch between the mechanical resonance frequencies will lift the EP degeneracy, impairing the performance 
of the proposed sensor. It could be interesting to extend the detection process studied in the previous section 
to a system composed of non-degenerated resonators. This can breakthrough the limitation of micro/nano-fabrication 
restriction, and will pave a way towards EP-sensors which are robust against defects or fabrication imperfections. 
One way to go beyond this technological limitation is to engineer nonlinearities, in order to control and compensate the 
frequency mismatch. Such optical control of frequency has been recently investigated to enhance energy transfer 
between two non-degenerated mechanical resonators \cite{[25],[26]}.
\begin{figure}[tbh]
\begin{center}
\resizebox{0.4\textwidth}{!}{
\includegraphics{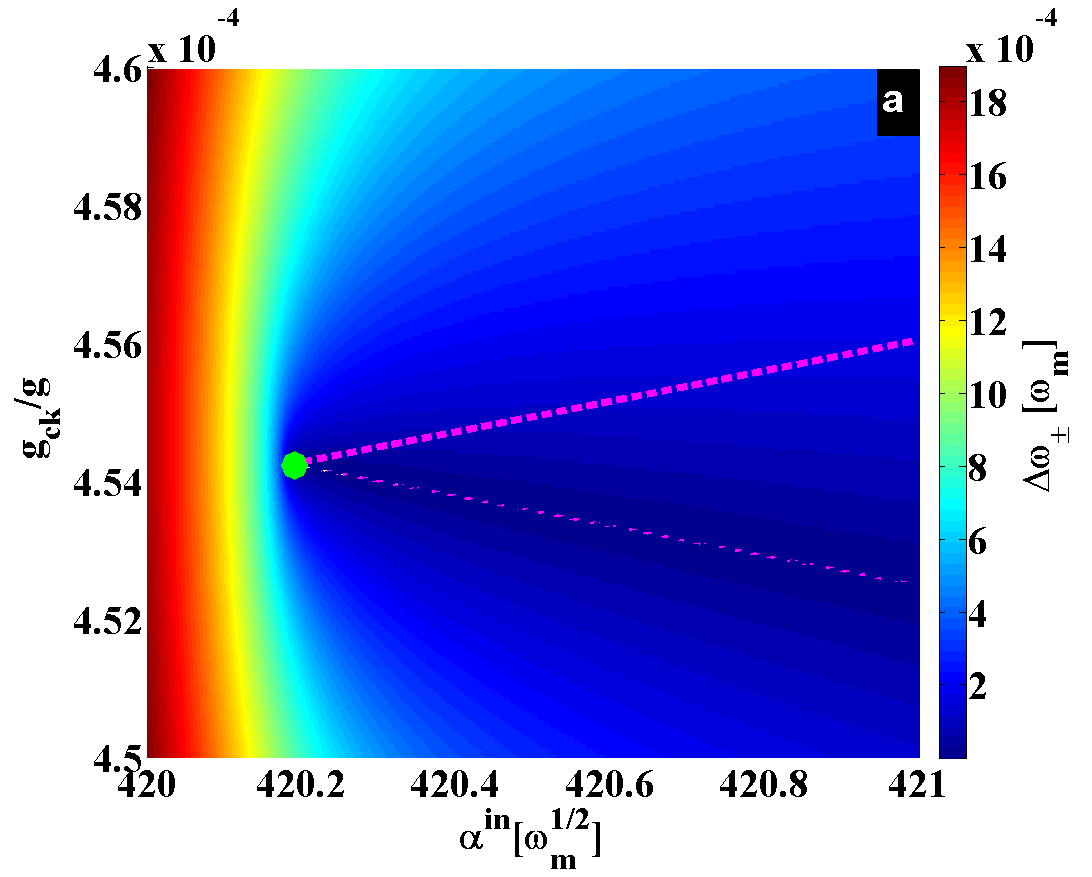}}
\resizebox{0.4\textwidth}{!}{
\includegraphics{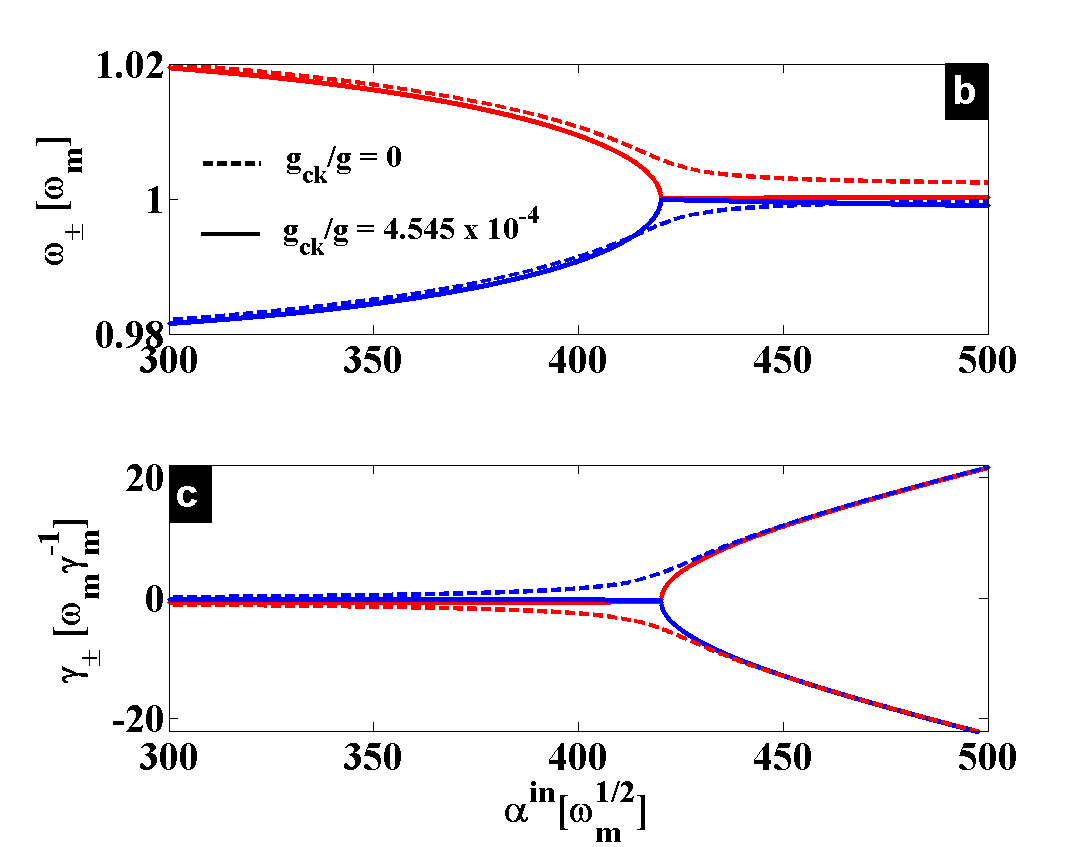}}
\end{center}
\caption{(a) Diagram showing the control strategy in ($\alpha^{in},g_{ck}$) space, at a fixed 
frequency mismatch $\nu=2\times 10^{-3}\omega_{m}$.  (b),(c) Eigenvalues corresponding to the green dot in (a), 
where we see EP feature induced by the control strategy. The dashed lines correspond to the case without 
control ($g_{ck}/g=0$), whereas the control has been applied ($g_{ck}/g=4.545\times 10^{-4}$) to the solid lines.}
\label{fig:Fig5}
\end{figure}
Duffing nonlinearity has been used in \cite{[25]}, whereas an optical trapping coupling was used in \cite{[26]} 
to trap the mechanical motion. Although these technics resulted to an efficient transfer 
around the avoided crossing level, this was still not enough to induce EP feature in the system. The reason lies on the fact that strong coupling observed in these works leads to level repulsion, 
gap opening between two modes due to coupling, while EP arises from level attraction process instead \cite{[27]}. 
To reach our goal, we propose a position-squared coupling where the optical mode is coupled in a 
quadratic fashion to the mechanical motion \cite{[21]}. This can be engineered by optically trapping 
the second resonator ($\beta_{2}$) inside the cavity driven on the blue-sideband,  which could have been 
achieved in \cite{[26]} as well. The resulting cross-Kerr interaction is described by the 
Hamiltonian $H_{ck}=-g_{ck}a^{\dag}ab^{\dag}b$, which has to be accounted for Eq. (\ref{eq1}) for a full 
corresponding model. After few transformations, the tuned mechanical frequency 
takes the form $\tilde{\omega_{2}}=\omega_{2}-g_{ck}\alpha_{2}^{\ast}\alpha_{2}$, which can be optically 
controlled as the cavity is driven (see appendix \ref{App.B}). By assuming that the coupled mechanical resonators fullfill 
the condition $\omega_{2} > \omega_{1}$, and for a given cross-Kerr term $g_{ck}>0$, one can optically
adjust the effective frequencies to satisfy $\omega_{\rm{eff}}^{1}\sim\omega_{\rm{eff}}^{2}$ that is a 
good requirement for EP as presented above. Therefore, cross-Kerr coupling is useful not only to 
engineer EP in non-degenerated coupled resonators, but it enables to enhance the performance of optomechanical 
sensor-based EP for mismatched set of resonators, regardless of their technological imperfections. 
Fig. \ref{fig:Fig5}(a) is a diagram showing the frequency control strategy in ($\alpha^{in},g_{ck}$) space, for a given 
mismatch frequency $\nu=\omega_{2} - \omega_{1}=2\times 10^{-3}\omega_{m}$. It can be seen that 
any mismatch ($\nu\neq0$) requires a corresponding cross-Kerr value to reach the minimal splitting 
$\Delta\omega_{\pm}$ (see colorbar in Fig. \ref{fig:Fig5}(a)), which is depicted by the magenta dashed line for a given
couple of parameters ($\alpha^{in},g_{ck}$). The eigenvalues that correspond to the green dot in Fig. \ref{fig:Fig5}(a) are represented in
Fig. \ref{fig:Fig5}(b), where we can see the efficiency of the strategy control as well as the EP feature. Indeed, the dashed lines
correspond to the case without control ($g_{ck}/g=0$), whereas the control has been applied ($g_{ck}/g=4.545\times 10^{-4}$) to the solid lines. 
The observed large splitting in the case without control simply reveals an absence of EP together with the fact that a small object 
can not be efficiently detected. Therefore, a position-squared coupling is a useful tool to tune non-degenerated mechanical 
resonators closer to their EP. Hence, detection of optomechanical sensor-based EP can be enhanced through quadratic coupling, 
resulting in a giant sensitivity factor compared to the conventional sensors.

\section{Conclusion} \label{Concl}
We have proposed an optomechanical mass sensor based on EP. The benchmark system is a mechanically
coupled optomechanical system, where gain (loss) is engineered by driving the cavity with a
blue (red) detuned electromagnetic field. Degeneracy known as EP shows up in the system when the gain 
balances losses, and any perturbation leads to a frequency splitting that scales as the square-root of the 
perturbation strength that greatly enhances the detection compared to the conventional optomechanical 
sensors. To breakthrough the limitation of micro/nanofabrication technology restriction, we have extended 
the performance of the proposed sensor to non-degenerated mechanical resonators by using
quadratic optomechanical coupling to tune the mismatch mechanical frequency. Therefore, 
our sensing  scheme is robust regardless of technological limitation related to the mechanical resonators 
ingineering, and can be applied to a versatile set of resonators. Moreover, this scheme does not require 
$\mathcal{PT}$-symmetric prerequisite, and can be extended to plethora of systems including, 
hybrid opto-electromechanical and superconducting microvawe resonators.

\section*{Acknowledgments}

This work was supported by the European Commission FET OPEN H2020
project PHENOMEN-Grant Agreement No. 713450.

\appendix \label{App}

\section{Calculation details related to the sensitivity and its enhancement factor} \label{App.A}
\begin{figure}[tbh]
\begin{center}
\resizebox{0.4\textwidth}{!}{
\includegraphics{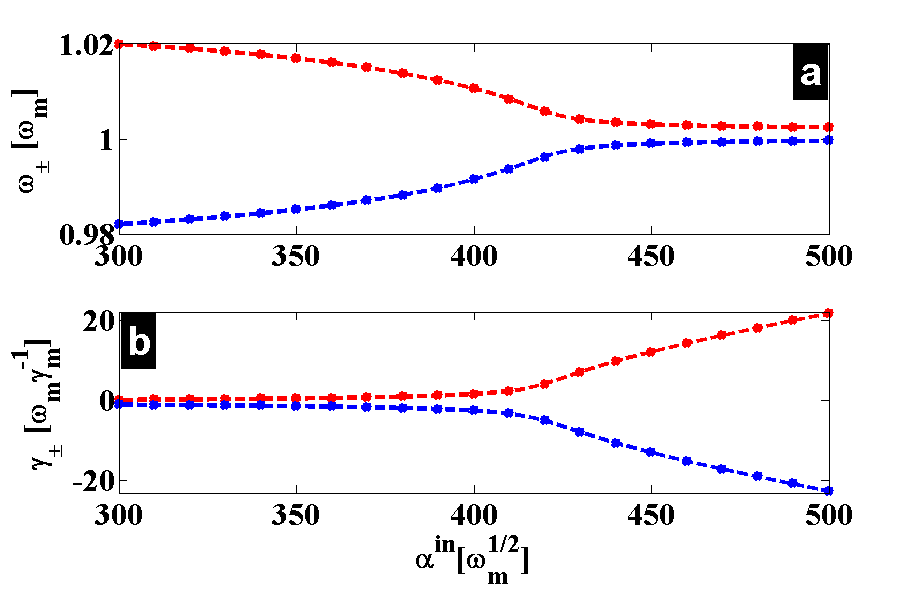}}
\resizebox{0.4\textwidth}{!}{
\includegraphics{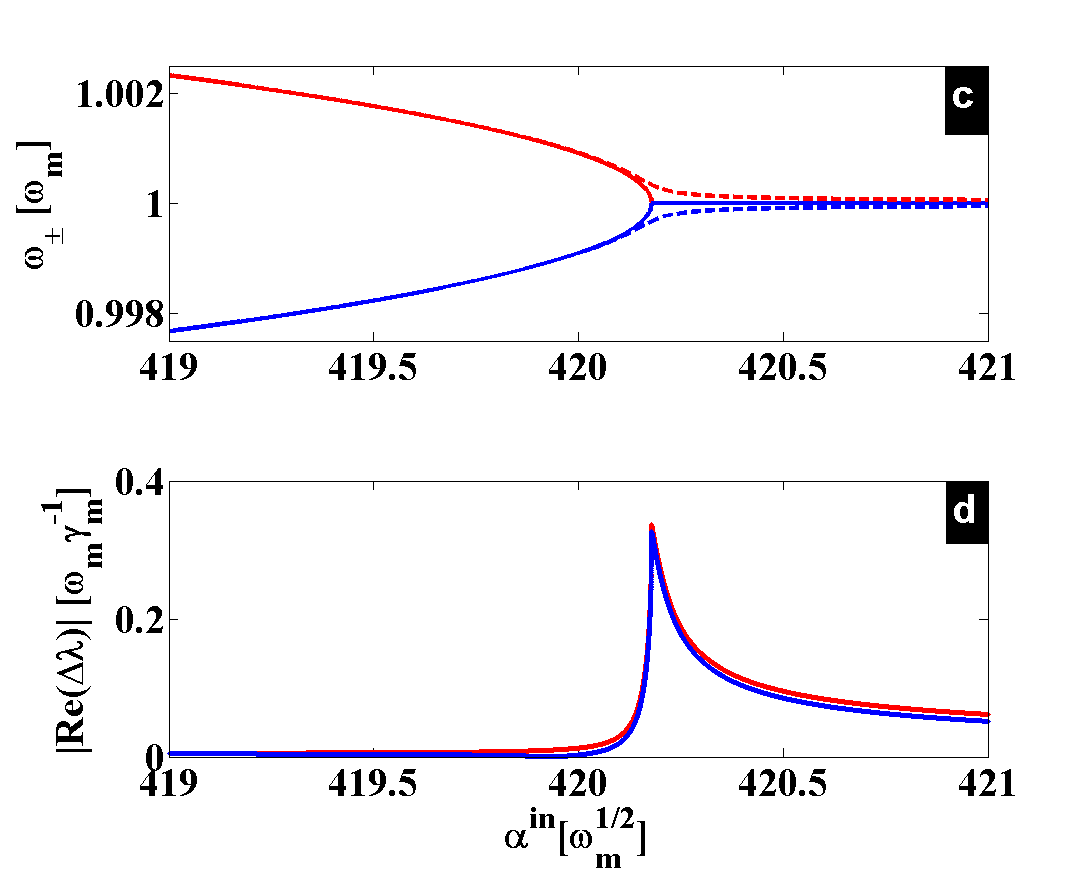}}
\end{center}
\caption{(a)-(b) Real and imaginary parts of the eigenvalues versus the driving field for a perturbation 
$\delta \omega=2\times 10^{-3}\omega_{m}$. The dashed lines are the numerical results, while the dotted lines 
represent the analytical approximation given in Eq. (\ref{eq11}) and Eq. (\ref{eqa5}). This 
validates the approximations made in our calculations. (c)-(d) Eigenfrequencies and the corresponding 
sensitivities for a weak perturbation $\delta \omega=10^{-5}\omega_{m}$. One remarks that both 
frequencies experience the same sensitivity as predicted in our calculations.}
\label{fig:Fig6}
\end{figure}

In the main text (see Eq. (\ref{eq5}) and Eq. (\ref{eq6})), we have given the eigenvalues of our effective reduced 
mechanical system as, 

\begin{equation}
\lambda_{\pm}=  \frac{\omega_{\rm{eff}}^{1}+\omega_{\rm{eff}}^{2}}{2}-\frac{i}{4}\left(\gamma_{\rm{eff}}^{1}+\gamma
_{\rm{eff}}^{2}\right) \pm \frac{\sigma}{4},  \label{eqa1}
\end{equation}
with 
\begin{equation}
\sigma = \sqrt{16J^{2}+[2(\omega_{eff}^{1}-\omega_{eff}^{2}) 
+i(\gamma_{eff}^{2}-\gamma_{eff}^{1})]^2}, \label{eqa2}
\end{equation}
where, $\omega_{\rm{eff}}^{j}=\omega_{m}+\delta \omega_{opt}^{j}$
and $\gamma_{\rm{eff}}^{j}=\gamma_{m}+\gamma_{\rm{opt}}^{j}$ are the effective frequencies 
and  dampings, respectively. Any perturbation ($\delta m$) of the second mechanical resonator 
(without loss to the generality) induces a frequency shift ($\delta \omega$) that affects the eigenvalues 
as, 
\begin{equation}
\lambda_{\pm}^{\delta \omega}=  \frac{\sum_{j=1,2} \omega_{\rm{j}}+\sum_{j=1,2} \delta\omega_{\rm{opt}}^{j}+\delta \omega}{2}-\frac{i}{4}\left(\gamma_{\rm{eff}}^{1}+\gamma
_{\rm{eff}}^{2}\right) \pm \frac{\sigma^{\delta \omega}}{4},  \label{eqa3}
\end{equation}
with 
\begin{equation}
\sigma^{\delta \omega} = \sqrt{16J^{2}+ \left[2\chi+i(\gamma_{eff}^{2}-\gamma_{eff}^{1})\right]^2}, \label{eqa4}
\end{equation}
where $\chi=\left(\omega_{\rm{1}}-\omega_{\rm{2}}+ \delta\omega_{\rm{opt}}^{1}-\delta\omega_{\rm{opt}}^{2}-\delta \omega\right)$, 
and we have localized the effect of $\delta \omega$ only on the frequencies, since it weakly affects the dampings 
(see Fig. \ref{fig:Fig6}(a)-(b)). However, the full effect is accounted in our numerical results shown in the main text. For degenerated mechanical 
resonators ($\omega_{j}\equiv\omega_{m}$), and in a weak driving regime ($\delta\omega_{\rm{opt}}^{j} \ll \omega_{m}$), 
Eq. (\ref{eqa3}) can be simplified as,
\begin{equation}
\begin{split}
\lambda_{\pm}^{\delta \omega} &= \omega_{m} + \frac{\delta \omega}{2}-\frac{i}{4}\left(\gamma_{\rm{eff}}^{1}+\gamma
_{\rm{eff}}^{2}\right)\\
& \pm \frac{1}{4}\sqrt{16J^{2}+ \left[-2\delta \omega+i(\gamma_{eff}^{2}-\gamma_{eff}^{1})\right]^2} \\
&=\omega_{m} + \frac{\delta \omega}{2}-\frac{i}{4}\left(\gamma_{\rm{eff}}^{1}+\gamma
_{\rm{eff}}^{2}\right)\\
&\pm \sqrt{J^{2}- \left(\frac{\Delta\gamma_{eff}}{4}\right)^{2}+\frac{\delta \omega^{2}
-i\delta \omega\Delta\gamma_{eff}}{4}}.  \label{eqa5}
\end{split}
\end{equation}

The sensitivity is measured by the detected change between the perturbed and the non-perturbed eigenvalues. From 
Eq. (\ref{eqa5}), one can deduce such quantities providing, 
\begin{equation}
\begin{split}
\lambda_{\pm}^{\delta \omega}-\lambda_{\pm} &=\frac{\delta \omega}{2} 
\pm \sqrt{J^{2}- \left(\frac{\Delta\gamma_{eff}}{4}\right)^{2}+\frac{\delta \omega^{2}
-i\delta \omega\Delta\gamma_{eff}}{4}} \\
& \mp \sqrt{J^{2}- \left(\frac{\Delta\gamma_{eff}}{4}\right)^{2}}.  \label{eqa6}
\end{split}
\end{equation}
As we seek to evaluate the sensitivity of our sensor at the EP where the quantity given in 
Eq. (\ref{eqa6}) has an optimal value, we use the condition $4J=\Delta\gamma_{eff}$ that leads to, 
\begin{equation}
\begin{split}
\lambda_{\pm}^{\rm{EP}}\equiv\left(\lambda_{\pm}^{\delta \omega}-\lambda_{\pm}\right)^{\rm{EP}}& =\frac{\delta \omega}{2} 
\pm \sqrt{\frac{\delta \omega^{2}-i\delta \omega\Delta\gamma_{eff}}{4}}\\
&=\frac{1}{2}\left(\delta \omega\pm \sqrt{\delta \omega^{2}-i\delta \omega\Delta\gamma_{eff}}\right).  \label{eqa7}
\end{split}
\end{equation}
In order to ease our qualitative interpretations of the physics hidden behind Eq. (\ref{eqa7}), it is 
convenient to explicitly split it in real and imaginary parts. Indeed, the square-root of the complex term 
in Eq. (\ref{eqa7}) can be found by first converting it to polar form and then 
using DeMoivre's theorem. After few calculations, we straightforwardly 
get,
\begin{equation}
\rm{Re}\left(\lambda_{\pm}^{\rm{EP}}\right)=\frac{1}{2}\left(\delta \omega \pm 
\sqrt{\frac{\delta\omega^{2}}{2}  + \frac{\delta\omega}{2} 
\sqrt{\delta \omega^{2}+\Delta\gamma_{eff}^{2}} }\right),  \label{eqa8}
\end{equation}
and 
\begin{equation}
\rm{Im}\left(\lambda_{\pm}^{\rm{EP}}\right)= \pm\frac{i}{2} 
\sqrt{\frac{\delta\omega}{2} 
\sqrt{\delta \omega^{2}+\Delta\gamma_{eff}^{2}} -\frac{\delta\omega^{2}}{2}}.  \label{eqa9}
\end{equation}
From Eqs. (\ref{eqa8})-(\ref{eqa9}), we can straightforwardly deduce the sensitivities 
$\left|\rm{Re}\left(\lambda_{\pm}^{\rm{EP}}\right)\right|$ and $\left|\rm{Im}\left(\lambda_{\pm}^{\rm{EP}}\right)\right|$, 
which are depicted in Fig. \ref{fig:Fig2}(b), (or Fig. \ref{fig:Fig3}(a)) and Fig. \ref{fig:Fig4}(a), 
respectively. Moreover, Eq. (\ref{eqa8}) reveals that the sensitivity given by the two eigenfrequencies 
are not equal in general, whereas the dampings carry out the same sensitivity through Eq. (\ref{eqa9}). 
Furthermore, Eqs. (\ref{eqa8})-(\ref{eqa9}) show that the sensitivity carried out through the dampings is generally less compared to the one 
coming from the eigenfrequencies ($\left|\rm{Im}\left(\lambda_{\pm}^{\rm{EP}}\right)\right|\leq 
\left|\rm{Re}\left(\lambda_{\pm}^{\rm{EP}}\right)\right|$). These qualitative explanations are revealed through 
Fig. \ref{fig:Fig2}(b) and Fig. \ref{fig:Fig4}(b). 

In a weak perturbation limit ($|\delta \omega|\ll|\Delta\gamma_{eff}|$), Eqs. (\ref{eqa8})-(\ref{eqa9}) are 
further simplified and carry out approximately the identical sensitivity through 
$\sim\pm\sqrt{\frac{\delta \omega\Delta\gamma_{eff}}{8}}$ and 
$\sim\pm i\sqrt{\frac{\delta \omega\Delta\gamma_{eff}}{8}}$, respectively.  In this limit, the sensitivity 
enhancement factor $\eta$ yields, 
\begin{equation}
\eta\equiv\left|\frac{\rm{Re}(\Delta\lambda_{\pm}^{\rm{EP}})}{\delta\omega}\right|=\sqrt{\frac{\Delta\gamma_{eff}}{8\delta \omega}}
=\sqrt{\frac{m\Delta\gamma_{eff}}{4\omega_{m} \delta m}}.  \label{eqa10}
\end{equation}
For a given weak perturbation $\delta \omega=10^{-5}\omega_{m}$, Figs. \ref{fig:Fig6}(c)-(d) show the 
eigenfrequencies and their corresponding sensitivities. It can be seen that the resulted sensitivities are 
similar for both eigenfrequencies as expected from our analysis. We would like to underline that in the main 
text, we have used $\lambda^{\rm{EP}}\equiv\lambda_{\pm}^{\rm{EP}}$ in order to simplify our notations.

\section{Frequency control through quadratic coupling} \label{App.B}
By accounting the quadratic coupling in the model, the Hamiltonian becomes,

\begin{equation}
H=H_{OM,ck}+H_{int}+H_{drive}, \label{eqb1}
\end{equation}
where $\rm{ck}$ stands for cross-Kerr, and
\begin{equation}
\left\{
\begin{array}
[c]{c}
H_{OM,ck}=\sum_{j=1,2}\omega_{j}b_{j}^{\dag}b_{j}-\Delta_{j} a_{j}^{\dag}a_{j}\\
-ga_{j}^{\dag}a_{j}(b_{j}^{\dag}+b_{j})-g_{ck}a^{\dag}ab^{\dag}b  \\
H_{int}=-J(b_{1}b_{2}^{\dag}+b_{1}^{\dag}b_{2})\\
H_{drive}=E(a^{\dag}+a).
\end{array}
\right. \label{eqb2}
\end{equation}
This leads to the following classical set of nonlinear equations,  
\begin{equation}
\left\{
\begin{array}{c}
\dot{\alpha}_{1}=[i(\Delta_{1}+g(\beta_{1}^{\ast}+\beta_{1})) 
-\frac{\kappa }{2}] \alpha_{1}-i\sqrt{\kappa}\alpha^{\rm{in}}, \\
\dot{\alpha}_{2}=[i(\Delta_{2}+g(\beta_{2}^{\ast}+\beta_{2})
+g_{ck}\beta_{2}^{\ast}\beta_{2}) -\frac{\kappa }{2}] \alpha_{2}-i\sqrt{\kappa}\alpha^{\rm{in}}, \\
\dot{\beta}_{1}=-(i\omega_{1}+\frac{\gamma_{m}}{2}) \beta_{1}+iJ\beta_{2}+ig\alpha_{1}^{\ast}\alpha_{1},\\
\dot{\beta}_{2}=-(i\tilde{\omega_{2}}+\frac{\gamma_{m}}{2}) \beta_{2}+iJ\beta_{1}+ig\alpha_{2}^{\ast}\alpha_{2},
\end{array}
\right.  \label{eqb3}
\end{equation}
where $\tilde{\omega_{2}} =\omega_{2}-g_{ck}\alpha_{2}^{\ast}\alpha_{2}$ is the optically tunable mechanical frequency mentioned 
in the main text.

\begin{figure}[tbh]
\begin{center}
\resizebox{0.5\textwidth}{!}{
\includegraphics{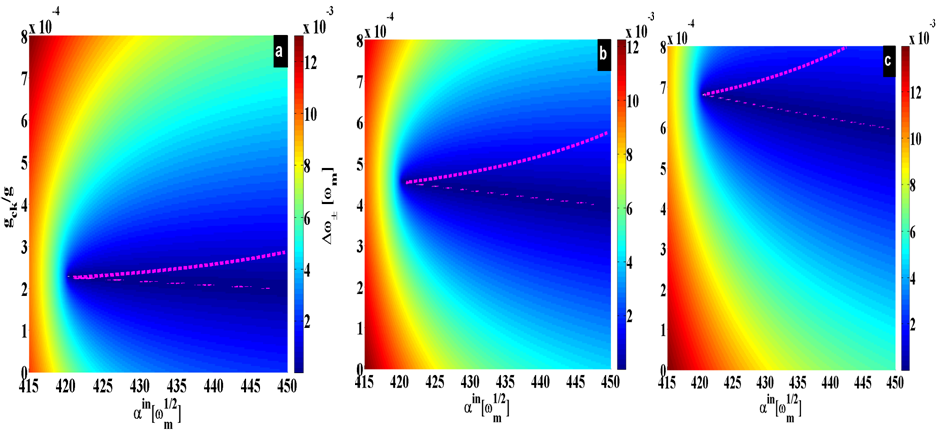}}
\end{center}
\caption{Diagrams showing the control strategy in ($\alpha^{in},g_{ck}$) space, at the 
frequency mismatches $\nu=1\times 10^{-3}\omega_{m}$, $\nu=2\times 10^{-3}\omega_{m}$ and $\nu=3\times 10^{-3}\omega_{m}$, 
for (a), (b), and (c) respectively. It is revealed that large mismatch frequency requires large 
cross-Kerr term to reach minimal splitting.}
\label{fig:Fig7}
\end{figure}

For an overview of the cross-Kerr effect, we have extended Fig. \ref{fig:Fig5}(a) of the main text to a large 
scale parameters space and for different frequency mismatch as shown in Fig. \ref{fig:Fig7}. One remarks that the 
cross-Kerr term required both to reach minimal splitting and to get closer to the EP increases as the mismatch frequency increases. This comes 
from the tuned frequency $\tilde{\omega_{2}} =\omega_{2}-g_{ck}\alpha_{2}^{\ast}\alpha_{2}$, which reveals the usefulness 
of cross-Kerr coupling not only in frequency control but also in sensitivity enhancement. 


\begin{thebibliography}{99}
\bibitem{[1]} M. Li, H. X. Tang, and M. L. Roukes, Ultra-sensitive NEMS-based 
cantilevers for sensing, scanned probe and very high-frequency applications, 
Nature Nanotech. \textbf{2}, 114 (2007).

\bibitem{[2]} K. L. Ekinci, X. M. H. Huang, and M. L. Roukes, 
Ultrasensitive nanoelectromechanical mass detection, Appl. Phys. Lett. \textbf{84}, 4469 (2004).

\bibitem{[3]} N. V. Lavrik and P. G. Datskos, Femtogram mass detection using 
photothermally actuated nanomechanical resonators, Appl. Phys. Lett. \textbf{82}, 2697 (2003).

\bibitem{[4]} B. Ilic, H. G. Craighead, S. Krylov, W. Senaratne, C. Ober, and P. Neuzil, 
Attogram detection using nanoelectromechanical oscillators,  J. Appl. Phys. \textbf{95}, 3694 (2004).

\bibitem{[5]} Y. T. Yang et \textit{al.}, Zeptogram-scale nanomechanical mass sensing,  
Nano Lett. \textbf{6}, 583 (2006).

\bibitem{[6]} J. Chaste et \textit{al.}, A nanomechanical mass sensor with yoctogram resolution, 
Nature Nanotech. \textbf{7}, 301 (2012).

\bibitem{[7]} J.-J. Li, and K.-D. Zhu, All-optical mass sensing with coupled mechanical resonator 
systems, Physics Reports \textbf{525}, 223 (2013).

\bibitem{[8]} J.-J. Li, and K.-D. Zhu, Nonlinear optical mass sensor with an optomechanical 
microresonator, Appl. Phys. Lett. \textbf{101}, 141905 (2012).

\bibitem{[9]} F. Liu et \textit{al.}, Sub-pg mass sensing and measurement with an optomechanical 
oscillator, Opt. Express \textbf{21}, 19555 (2013).

\bibitem{[10]} H. Xiong, L.-G. Si, and Y. Wu, Precision measurement of electrical 
charges in an optomechanical system beyond linearized dynamics, Appl. Phys. Lett. \textbf{110}, 171102 (2017).

\bibitem{[11]} C. Jiang et \textit{al.}, Ultrasensitive nanomechanical mass sensor using 
hybrid opto-electromechanical systems, Opt. Express \textbf{21}, 13773 (2014).

\bibitem{[12]} Y. He, Sensitivity of optical mass sensor enhanced by 
optomechanical coupling, Appl. Phys. Lett. \textbf{106}, 121905 (2015).

\bibitem{[13]} J. Hakansson \textit{al.}, Strong forces in optomechanically
actuated resonant mass sensor, Opt. Express \textbf{25}, 30939 (2017).

\bibitem{[14]} I. M. Haghighi et \textit{al.}, Sensitivity-Bandwidth 
Limit in a Multimode Optoelectromechanical Transducer, Phys. Rev. Applied \textbf{9}, 034031 (2018).

\bibitem{[15]} T. S. Biswas et \textit{al.}, Time-Resolved Mass Sensing of a Molecular 
Adsorbate Nonuniformly Distributed Along a Nanomechnical String Phys. Rev. Applied \textbf{3}, 064002 (2015).

\bibitem{[16]} J. Wiersig, Enhancing the sensitivity of frequency and energy splitting detection by using 
exceptional points: Application to microcavity sensors for single-particle detection, Phys. Rev. Lett. \textbf{112}, 203901 (2014).

\bibitem{[17]} W. Chen et \textit{al.}, Exceptional points enhance sensing in an optical 
microcavity, Nature \textbf{548}, 192 (2017); Parity-time-symmetric whispering-gallery mode 
nanoparticle sensor, Photonics Research \textbf{6}, A23 (2018).

\bibitem{[18]} H.  Hodaei et \textit{al.}, Enhanced sensitivity at higher-order 
exceptional points, Nature \textbf{548}, 187 (2017).

\bibitem{[19]} Z.-P. Liu et \textit{al.}, Metrology with PT-Symmetric Cavities: 
Enhanced Sensitivity near the $\mathcal{PT}$-Phase Transition, Phys. Rev. Lett. \textbf{117}, 110802 (2016).

\bibitem{[20]} P. Djorwe, Y. Pennec, and B. Djafari-Rouhani, Frequency locking and controllable 
chaos through exceptional points in optomechanics, Phys. Rev. E \textbf{98}, 032201 (2018).

\bibitem{[21]} T. K. Paraïso et \textit{al.}, Position-Squared Coupling in a Tunable Photonic Crystal 
Optomechanical Cavity, Phys. Rev. X \textbf{5}, 041024 (2015).

\bibitem{[22]} J. D. Cohen et \textit{al.}, Phonon counting and intensity interferometry 
of a nanomechanical resonator, Nature \textbf{520}, 522 (2015).

\bibitem{[23]} S. Hong et \textit{al.}, Hanbury Brown and Twiss interferometry of single 
phonons from an optomechanical resonator, Science \textbf{358}, 203 (2017).

\bibitem{[24]} H. Xu et \textit{al.}, Topological energy transfer in an optomechanical system 
with exceptional points, Nature \textbf{537}, 80 (2016).

\bibitem{[25]} M. Pernpeintner et \textit{al.}, Frequency control and coherent excitation transfer in a nanostring-resonator 
network, Phys. Rev. Applied \textbf{10}, 034007 (2018).

\bibitem{[26]} H. Fu et \textit{al.}, Coherent optomechanical switch for motion transduction based on dynamically localized 
mechanical modes, Phys. Rev. Applied \textbf{9}, 054024 (2018).

\bibitem{[27]} N. R. Bernier et \textit{al.}, Level attraction in microwave optomechanical circuit, Phys. Rev. A \textbf{98}, 
023841 (2018).


\end{thebibliography}
\end{document}